\documentclass[iop,revtex4]{emulateapj}
\newcommand{\mytilde}{\raise.19ex\hbox{$\scriptstyle\sim$}}

\newcommand\cluster{CL2023}
\usepackage{amsmath}
\usepackage{xcolor}
\usepackage{soul}
\usepackage{upgreek}
\graphicspath{{./}{figures/}}

\shorttitle{SPT-CL 2023-5535}
\shortauthors{HyeongHan et al.}

\begin{document}

\title{Discovery of a Radio Relic in the Massive Merging Cluster SPT-CL~2023-5535 from the ASKAP-EMU PILOT SURVEY}

\author{Kim HyeongHan\altaffilmark{1}, M. James Jee\altaffilmark{1,2},
Lawrence Rudnick\altaffilmark{3},
David Parkinson\altaffilmark{4}, 
Kyle Finner\altaffilmark{1}, 
Mijin Yoon\altaffilmark{1,5},
Wonki Lee\altaffilmark{1},
Gianfranco Brunetti\altaffilmark{6},
Marcus Br\"{u}ggen\altaffilmark{7},
Jordan D. Collier\altaffilmark{8,9}, 
Andrew M. Hopkins\altaffilmark{10},
Micha{\l}~J.~Micha{\l}owski\altaffilmark{11},
Ray P. Norris\altaffilmark{12,13},
Chris Riseley\altaffilmark{14,15,16} 
}

\altaffiltext{1}{Yonsei University, Department of Astronomy, Seoul, Republic of Korea; ahrtears54@yonsei.ac.kr, mkjee@yonsei.ac.kr}
\altaffiltext{2}{Department of Physics, University of California, Davis, California, USA}
\altaffiltext{3}{Minnesota Institute for Astrophysics, University of Minnesota, Minneapolis, Minnesota, USA}
\altaffiltext{4}{Korea Astronomy and Space Science Institute 776, Daedeokdae-ro, Yuseong-gu, Daejeon, Republic of Korea}
\altaffiltext{5}{Ruhr-University Bochum, Astronomical Institute, German Centre for Cosmological Lensing, Universitätsstr. 150, 44801 Bochum, Germany}
\altaffiltext{6}{Istituto Nazionale di Astrofisica, Istituto di Radioastronomia Via P Gobetti 101, 40129 Bologna, Italy}
\altaffiltext{7}{Hamburger Sternwarte, Universit\"at Hamburg, Gojenbergsweg 112, 21029 Hamburg, Germany}
\altaffiltext{8}{Inter-University Institute for Data Intensive Astronomy, Department of Astronomy, University of Cape Town, Private Bag X3, Rondebosch, 7701, South Africa}
\altaffiltext{9}{School of Science, Western Sydney University, Locked Bag 1797, Penrith, NSW 2751, Australia}
\altaffiltext{10}{Australian Astronomical Optics, Macquarie University, 105 Delhi Rd, North Ryde, NSW 2113, Australia}
\altaffiltext{11}{Astronomical Observatory Institute, Faculty of Physics, Adam Mickiewicz University, ul.~S{\l}oneczna 36, 60-286 Pozna{\'n}, Poland}
\altaffiltext{12}{Western Sydney University, Locked Bag 1797, Penrith South, NSW 1797, Australia}
\altaffiltext{13}{CSIRO Astronomy \& Space Science, PO Box 76, Epping, NSW 1710, Australia}
\altaffiltext{14}{Dipartimento di Fisica e Astronomia, Universit\`a degli Studi di Bologna, via P. Gobetti 93/2, 40129 Bologna, Italy}
\altaffiltext{15}{INAF -- Istituto di Radioastronomia, via P. Gobetti 101, 40129 Bologna, Italy}
\altaffiltext{16}{CSIRO Astronomy and Space Science, PO Box 1130, Bentley, WA 6102, Australia}

\begin{abstract}
The ASKAP-EMU survey is a deep wide-field radio continuum survey designed to cover the entire southern sky and a significant fraction of the northern sky up to $+30^{\circ}$. 
Here, we report a discovery of a radio relic in the merging cluster SPT-CL 2023-5535 at $z=0.23$ from the ASKAP-EMU pilot 300 sq. deg survey (800$-$1088~MHz). 
The deep high-resolution data reveal a $\mytilde2$ Mpc-scale radio halo elongated in the east-west direction, coincident with the intracluster gas. 
The radio relic is located at the western edge of this radio halo stretched $\mytilde0.5$~Mpc in the north-south orientation.
The integrated spectral index of the radio relic within the narrow bandwidth is $\alpha^{\scriptstyle \rm 1088~MHz}_{\scriptstyle \rm 800~MHz}=-0.76 \pm 0.06$. 
Our weak-lensing analysis shows that the system is massive ($M_{200}=1.04\pm0.36\times 10^{15}~M_{\sun}$) and composed of at least three subclusters. 
We suggest a scenario, wherein the radio features arise from the collision between the eastern and middle subclusters.
Our discovery illustrates the effectiveness of the ASKAP-EMU survey in detecting diffuse emissions in galaxy clusters and when completed, the survey will greatly increase the number of merging cluster detections with diffuse radio emissions.

\end{abstract}
\keywords{galaxies: clusters : 
individual (SPT-CL 2023-5535), gravitational weak lensing, radio continuum: radio relic and halo, X-rays:galaxies:clusters}

\section{Introduction} \label{sec:intro}
Large-scale diffuse radio emissions provide critical information for understanding galaxy cluster mergers.
They can be classified into two broad categories: halos and relics.
Radio halos are diffuse sources without distinct optical counterparts and are found in the central regions of merging clusters. Radio relics also do not have optical counterparts, but they are located in the cluster periphery and in general possess high levels of polarization.
Although both radio halos and relics are  indicators of cluster merger activities, radio relics have been considered a stronger constraint on the merger history because they can be used as direct probes of merger shocks \citep[see reviews for][]{2008ferrari, 2012feretti, 2019vanweeren}.

The kinetic energy that is dissipated during cluster-cluster mergers can power the observed cluster-scale radio emission. However, the complex chain of physical mechanisms that leads to the acceleration of emitting particles and amplification of magnetic fields in the ICM are still poorly understood \citep[e.g.,][]{2014brunetti}.
Giant radio halos are thought to originate from stochastic re-acceleration induced by cluster mergers turbulence \citep[e.g.,][]{2001brunetti, 2001petrosian, 2007brunetti, 2015miniati}; the contribution from secondary particles generated by the chain of hadronic collisions in the ICM has also been explored in the past \citep[e.g.,][]{1980dennison, 1999blasi} and more recently in combination with turbulent reacceleration models \citep[e.g.,][]{2011brunetti, 2017pinzke}.
Radio relics are believed to originate from merger shocks \citep[e.g.,][]{1998ensslin}.
The original approach was based on the diffusive shock acceleration \citep[DSA;][]{1978bell, 1983drury, 2001malkov} of thermal electrons. 
However, the efficiency of acceleration at weak shocks in the ICM appears too low to reproduce the spectrum and luminosity of a large fraction of the observed radio relics \citep[e.g.,][]{2020botteon}.
One of the most popular modifications of this scenario is based on shock re-acceleration of pre-existing relativistic plasma \citep[e.g.,][]{2011kang, 2013pinzke, 2016kang}, which has been supported in some cases by the connection between radio relics and AGNs \citep[e.g.,][]{2014bonafede, 2017vanweeren}.

In order to better understand the origin of radio halos and relics, a current priority should be to increase the sample size.
To date, there are $\mytilde60$ known radio relics.
Because of the cluster-to-cluster variation, the existing sample is too small to enable studies, where one can extract overarching principles.
Among the upcoming concerted efforts, presently the Australian Square Kilometre Array Pathfinder\footnote{https://www.atnf.csiro.au/projects/askap/index.html}-Evolutionary Map of the Universe\footnote{https://www.emu-survey.org/} \citep[ASKAP-EMU;][]{2011norris} is the largest deep ($\mytilde10~\mu\mbox{Jy/beam}$), high-resolution ($\mytilde10\arcsec$) radio continuum survey designed to cover the entire southern sky and a significant fraction of the northern sky up to $+30^{\circ}$. 
One of the scientific goals of the project is to enlarge the sample of clusters with diffuse radio emissions by at least two orders of magnitude.

In this study, we report the discovery of a radio relic in the massive merging cluster SPT-CL 2023-5535 (hereafter \cluster~for brevity) at $z=0.23$.
The presence of the diffuse radio emission in \cluster~has been reported in Zheng et al. (in prep.), who used the Murchison Widefield Array\footnote{http://www.mwatelescope.org/} \citep[MWA;][]{2013tingay}, the Australia Telescope Compact Array\footnote{https://www.narrabri.atnf.csiro.au/} \citep[ATCA;][]{1992frater}, and the Molonglo Observatory Synthesis Telescope\footnote{https://astronomy.swin.edu.au/research/utmost/?page$\char`_$id=32} \citep[MOST;][]{1981mills, 1991robertson} data.
However, the insufficient spatial resolution and the several bright neighboring radio point sources have prevented the earlier work from clearly resolving the halo and relic. 
In this paper, we also present our weak-lensing (WL) and X-ray analyses of \cluster~based on the archival Dark Energy Camera \citep[DECam;][]{2015flaugher}  and {\it Chandra} data, respectively, which enhance interpretation of the current discovery.

We adopt a $\Lambda$CDM cosmology with $H_0=70$ km~s$^{-1}$Mpc$^{-1}$, $\Omega_m=0.3$, and $\Omega_{\Lambda}=0.7$. The angular size of $1\arcmin$ corresponds to a length scale of $\mytilde223$~kpc at the cluster redshift $z=0.23$. 

\section{Observations} \label{sec:observations}
\subsection{Radio} \label{sec:radio}

ASKAP has 36 antennae, 34 of which are placed within a region of ~2.3 km diameter while the outer four extend the baselines up to \mytilde6 km. In all cases, as many as 36 antennae were used, and always the 4 outer antennae were included because their uv coverage was critical to the imaging quality. However, in a few cases, some of the inner antennae were omitted because of maintenance or hardware issues.

At the focus of each antenna is a phased array feed (PAF), which subtends a solid angle of \mytilde30 sq. degrees. Each PAF consists of 192 dual-polarization receivers. A weighted sum of the outputs of groups of receivers form 30 beams. Individual receivers, in general, contribute to more than one beam. Therefore, adjacent beams are not completely independent. The 30 beams together cover an area of \mytilde30 sq. degrees on the sky. 
				
The weights of the individual beams are initially calibrated by observing the Sun placed successively at the center of each beam, and then adjusting the weights for maximum signal-to-noise. However, a radiator at the vertex of each antenna or the On-Dish Calibrator (ODC) enables the gain of each receiver to be calibrated. As a result, the solution initially obtained from the Sun observation is modified using the ODC calibration and used to adjust the weights.
				
Before (or sometimes after) the observation of each target, the calibrator source 1934-638 is observed for 200~s at the center of each of the 30 beams, to provide bandpass and gain calibration. No further calibration, other than self-calibration, is performed during the observation.

The radio data used in this paper were taken in 2019 July from the ASKAP-EMU  Pilot Survey, based on 10~h integration (a rms noise level of $25-35\rm~\mu$Jy/beam) for Scheduling Block 9351, with a frequency range of 800-1088 MHz.
The reduction was performed with the {\tt ASKAPsoft}\footnote{https://www.atnf.csiro.au/computing/software/askapsoft/sdp/\\docs/current/index.html} pipeline, using a multi-scale CLEAN algorithm and two Taylor terms (T0 and T1), which allow production of maps at a fiducial frequency of 943 MHz (T0) and the corresponding spectral indices (T1/T0).  
A more extensive description of the Pilot Survey will be provided by Norris et al. (in prep.).

In this paper we present images from both the original  (Figure~\ref{fig1}A) and diffuse-enhanced (Figure~\ref{fig1}B) versions. 
The latter was created by first masking out bright ($>40$ mJy) compact sources from the original version, then smoothing the masked image with a FWHM=25\arcsec~Gaussian kernel, and finally combining the smoothed image (purple) with the original image (green). 
The resulting image (Figure~\ref{fig1}B) makes it easy to visually separate diffuse emissions from compact sources.

\begin{table}
	\centering
	\caption{DECam observation}
	\label{obs}
	\begin{tabular}{ccccc} 
	    \hline
	    \hline
		 Filter & Date & $t_{exp}$ & Seeing  & $m_{lim}^1$\\
		        &      & (s)       & (arcsec)&          \\ 
		\hline
	    {\it g} & 2016 July 7 & 4875 & 1.46 & 26.0\\
		{\it r} & 2016 July 7 & 3375 & 1.33 & 25.8\\
		{\it i} & 2016 July 6 & 3625 & 0.78 & 25.0\\
        \hline
        \\
	\end{tabular}
\tablecomments{1. This is the 5$\sigma$ limiting magnitude for point sources.}
\end{table}

\subsection{Optical}

\cluster~was observed with the DECam mounted on the 4-meter Blanco telescope at the Cerro Tololo Inter-American Observatory (PI: von der Linden).
Table~\ref{obs} summarizes the observations for the {\it g}, {\it r}, and {\it i} filters that we retrieved from the NOAO archive\footnote{http://archive1.dm.noao.edu/} for the current study.
The Community Pipeline \citep{2014valdes} is used for the basic data reduction (i.e., overscan, bias, flat, etc.).
The calibrated images were stacked into a single mosaic image for each filter using $\tt{SCAMP}$\footnote{https://www.astromatic.net/software/scamp} and $\tt{SWARP}$\footnote{https://www.astromatic.net/software/swarp}.
We used the {\it i}-band image for our WL analysis because it provides the sharpest point spread function (PSF).
Intermediate PSF models were constructed for the individual exposures through principal component analysis (PCA; Jee et al. 2007) and stacked to obtain the final PSF model for shape measurement. 
Readers are referred to the descriptions in our previous papers for detail \citep[e.g.,][]{2011jee, 2013jee, 2017finner}.
After applying our S/N, color, magnitude, and shape measurement error cuts, we obtain a source density of $\mytilde11$ galaxies per sq. arcmin.

\begin{table}
	\centering
	\caption{Diffuse Radio Emission Properties}
	\label{table_radio}
	\begin{tabular}{lcc} 
		\hline
		\hline
		         & Halo        & Relic  \\
		\hline
	    $S_{943 \rm MHz}$ (mJy) & 31.3$\pm$0.6 & 16.2$\pm$0.2  \\
	    $S_{1.4 \rm GHz}$ (mJy)$^1$ & 20.8$\pm$0.3 & 12.0$\pm$0.3  \\
	    $P_{1.4 \rm GHz}$ ($10^{24}~\rm W~Hz^{-1}$) & 3.4 $\pm$ 0.1 & 1.8$\pm$0.1  \\
		Spectral Index ($\alpha$)           &  -1.04$\pm$0.05 & -0.76$\pm$0.06 \\
        \hline
	\end{tabular}
\tablecomments{1. Flux densities at 1.4 GHz are extrapolated assuming a power law.}
\end{table}

\subsection{X-ray}
\cluster~was observed with the {\it Chandra} X-ray observatory on 2014 March 30 (ObsId: 15108 - PI: Jones, ACIS-I detector, VFaint Mode, 20.81~ks).
The data were reduced using the $\tt{CIAO}$ $\tt{4.11}$ software with $\tt{CALDB}$ $\tt{4.8.3}$.
We reprocessed the raw data using the $\tt{chandra\char`_repro}$ script to produce a level 2 event file.
We then created a broad band (0.5-7~keV) exposure-corrected image with the $\tt{fluximage}$ script.

For our X-ray temperature measurement, point sources were masked out using the $\tt{wavdetect}$ script and background flares were removed with the $\tt{deflare}$ script.
We extracted grouped X-ray spectra with the $\tt{specextract}$ script in such a way that each bin has a minimum signal-to-noise ratio of 5.
Then, we performed spectral fitting with the $\tt{XSPEC}$ ($\tt{v12.10.1f}$) package and used the absorbed $\tt{MEKAL}$ plasma model \citep[][]{1993kaastra, 1995liedahl} within the 1-5~keV energy band.
The Galactic hydrogen density and the cluster metal abundance were assumed to be $N_H=5.1 \times 10^{20} \rm ~cm^{-2}$ \citep[][]{1990dickey} and 0.3 solar, respectively.

\begin{figure*}
	\includegraphics[width=2\columnwidth]{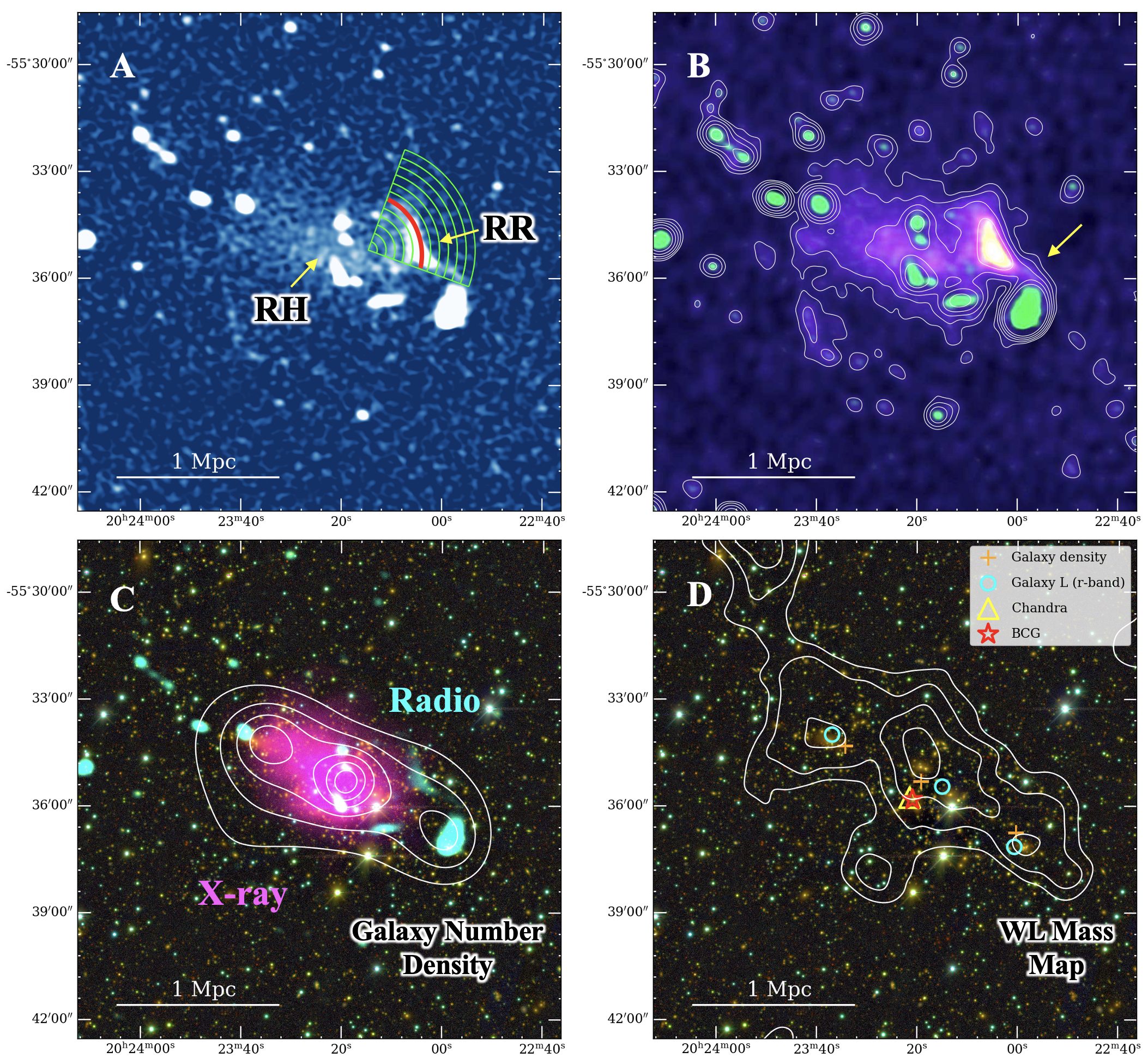}
    \caption{Multi-wavelength observations of \cluster.
    (A) ASKAP-EMU full-resolution ($12\arcsec \times11\arcsec$) image. 
    The green pan-shape regions indicate the areas for the X-ray surface brightness analysis. 
    The approximate location of the leading edge of the relic is marked with the red arc. 
    (B) Composite radio image after both diffuse (purple) and compact sources (green) are enhanced (\textsection\ref{sec:radio}). The diffuse radio halo is elongated $\mytilde1$~Mpc in the east-west orientation and at the western edge of the halo is the radio relic. White contours denote the $3\sigma_{rms} \times 2^{n}$ levels of the smoothed (FWHM=25\arcsec) image where $n=0, 1, 2, 3, 4$ and $\sigma_{rms}=\mytilde10~\mu\mbox{Jy/beam}$. The yellow arrow points at the ``link" feature between the relic and nearby radio galaxy (\textsection\ref{sec:discussion}).
    (C) Optical color composite image with the overlays of ASKAP-EMU radio (cyan), {\it Chandra} X-ray (magenta), and galaxy iso-density contours (white). We use the DECam {\it g}, {\it r}, and {\it i} filters to represent the intensities in blue, green, and red, respectively. A total of $\mytilde1800$ cluster member candidates are selected based on their 4000\AA~break features. 
    (D) Same as (C) except that we overlay the WL mass contours (white) over the optical color-composite image. The mass distribution composed of three substructures is consistent with the galaxy density distribution. The spectroscopically confirmed BCG coincides with the X-ray peak.}
    \label{fig1}
\end{figure*}

\section{Results} \label{sec:result}
\subsection{Detection of a Radio Relic and Halo} \label{sec:radio_results}
Cross-matching the galaxy clusters detected in the Planck SZ survey \citep[][]{2016planck} with the ASKAP-EMU radio continuum survey, we discovered a $\mytilde0.5$~Mpc relic associated with \cluster~shown in Figure~\ref{fig1}, based on its spectral index, morphology, orientation, and location.
Also, a clear $\mytilde1$~Mpc radio halo is detected eastward of the relic, which was originally reported by Zheng et al. (in prep.) as a halo candidate, whose observations however could not separate the halo from the relic because of the insufficient resolution.

The halo follows the diffuse X-ray emission (Figure~\ref{fig1}C) whose peak coincides with the BCG (Figure~\ref{fig1}D). 
The similarity in morphology between radio halo and ICM is typical of radio halos \citep[e.g.,][for a review]{2001govoni, 2001feretti, 2019vanweeren}. 
In the context of turbulent re-acceleration models the non-thermal components (particles and magnetic fields) are powered by the damping of the energy flux of large scale turbulence that is generated by the dynamics and energy density of the ICM component leading to a morphological connection between the two components \citep[e.g.,][for a review]{2014brunetti}; the details of the spatial correlation are sensitive to the way turbulence is generated in the thermal background plasma and relativistic particles are accelerated and transported in that turbulence.

\begin{figure}[t]
	\includegraphics[width=1\columnwidth]{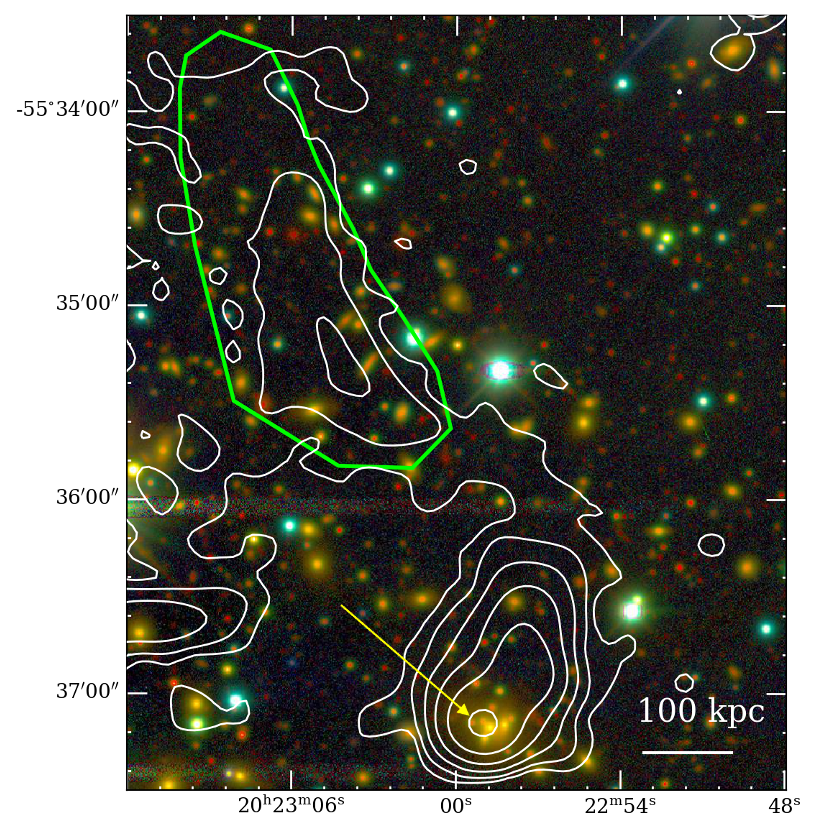}
    \caption{Close-up view of the western radio galaxy mentioned in \textsection\ref{sec:radio_results}. The optical counterpart is indicated by a yellow arrow whose radio lobe is connected to the relic. Green polygon defines the area of the relic for the flux measurement. White contours denote the $3\sigma_{rms} \times 3^{n}$ levels of the full-resolution ($12\arcsec\times11\arcsec$) image where $n=0, 1, ..., 5, 6$.}
    \label{radio_galaxy}
\end{figure}

The relic, elongated in the north-south orientation, lies $\mytilde0.5$~Mpc west of the BCG.
This location also corresponds to the western edge of the diffuse X-ray emission. 
Close inspection of the relic reveals that its southern edge is connected to a compact radio source via a faint ``link" (yellow arrow in Figure~\ref{fig1}B) reminiscent of the feature in Abell 3411-12 \citep[][]{2017vanweeren}.
The compact radio source has an optical counterpart (Figure~\ref{radio_galaxy}), which has a consistent color with that of the cluster red sequence. 
The location of this radio source also coincides with the western WL mass peak (see \textsection\ref{sec:weak_lensing}).

\begin{figure}[t]
	\includegraphics[width=1\columnwidth]{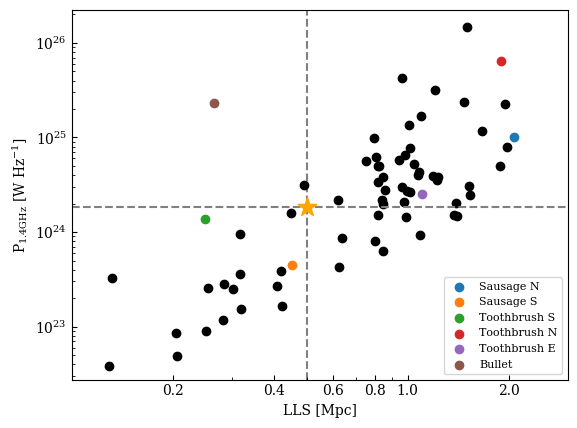}
    \caption{Monochromatic luminosity of relics at 1.4 GHz compared to their largest linear size (LLS) retrieved from \cite{2019vanweeren}. Orange star indicates the relic discovered in \cluster.}
    \label{LLS}
\end{figure}

We determined the radio flux density of the relic to be $S_{943 \rm MHz}=16.2\pm0.2$~mJy from the T0 image using a polygon aperture (see Figure~\ref{radio_galaxy}) that traces the visual boundary of the feature. Dividing the T1 image by the T0 image gives an integrated spectral index of $\alpha_{int}=-0.76 \pm 0.06$. The extrapolated radio flux density of the relic at 1.4~GHz is $S_{1.4 \rm GHz}=12.0 \pm 0.3$~mJy under the assumption that the spectral feature follows a single power law.
The monochromatic luminosity and largest linear size (LLS) of \cluster~are compared with those of other clusters retrieved from \cite{2019vanweeren} in Figure~\ref{LLS}, which shows that \cluster~follows the relation.

To measure the radio flux density of the halo, we first masked out bright point sources ($>40$~mJy) and then replaced the fluxes with the in-halo average value.
Within a polygon enclosing the halo, the flux density is $S_{943 \rm MHz}=31.3\pm0.6$ mJy. The integrated spectral index of the halo and the extrapolated radio flux density at 1.4~GHz are $\alpha_{int}=-1.04\pm0.05$ and
$S_{1.4 \rm GHz}= 20.8\pm0.3$~mJy, respectively, when we follow the same procedures used for the relic.
Note that here we only quote statistical errors based on our rms noise measurement. The total errors should be larger when also systematic errors (e.g., flux scaling errors) are included. Full characterization of the systematic errors will be progressing over the next few years.
We summarize the above radio property measurements in Table~\ref{table_radio}.

\subsection{Cluster Galaxy and Weak-lensing Mass Distribution} \label{sec:weak_lensing}

Detection of the radio relic suggests that \cluster~underwent a major merger. 
However, in order to reconstruct the merger scenario, we need to identify the cluster substructures contributing to the merger. 
To this end, we use both the galaxy and mass distributions. 

Because no spectroscopic data of \cluster~are publicly available, we selected the cluster member candidates based on their 4000\AA~break features. 
From the color-magnitude diagram, we chose a total of $\mytilde1800$ member candidates within the $\mytilde 13\arcmin\times13\arcmin$ region approximately centered at the BCG.
We adaptively smoothed the galaxy number density using the {\tt csmooth} tool with a minimum significance of 2.5$\sigma$. 
The resulting iso-density contours are displayed in Figure~\ref{fig1}C. 
The galaxy distribution suggests that \cluster~consists of three subclusters distributed in the east-west orientation. 
The central component is the most significant clump, coincident with the X-ray and radio halo centroids.

This three-component structure seen in the cluster galaxy distribution is in excellent agreement with our WL mass distribution. 
We show the mass map obtained with the {\tt FIATMAP} \citep[][]{1997fischer} code in Figure~\ref{fig1}D, and we verified that very similar mass distributions are obtained with different algorithms such as the {\tt MAXENT} \citep[][]{2007jee} or 
Fourier-inversion \citep[][]{1993kaiser} methods.
Using bootstrapping analysis, we estimate that the central mass peak has the highest significance ($5.0\sigma$), followed by the eastern ($3.5\sigma$) and then by the western ($3.0\sigma$) peaks.
By fitting three Navarro-Frenk-White \citep[NFW;][]{1997navarro} profiles simultaneously, we find that the eastern, central, and western masses are $M_{200c}=2.6\pm1.6\times10^{14}M_{\sun}$, $3.5\pm1.7\times10^{14}M_{\sun}$, and $1.5\pm1.2\times10^{14}M_{\sun}$, respectively (Table~\ref{tab:mass}). Under the assumption that the three clumps are at the same distance from us, the total mass of the system is estimated to be $M_{200c}=1.04\pm0.36\times10^{15}~M_{\sun}$\footnote{The total mass is greater than the sum of the three substructures because $r_{200}$ also increases.}.

\begin{table}
    \caption{Mass Estimates of Substructures from WL analysis} 

    \centering
	\label{tab:mass}
	\begin{tabular}{ccc}
	   \hline
	   \hline
       Substructure &
       $M_{200c}^1$ & Peak Significance$^2$\\
         &
       ($10^{14} M_{\sun}$) & ($\sigma$) \\
		\hline
	    East & 2.6 $\pm$ 1.6  & 3.6 \\
		Center & 3.5 $\pm$ 1.7 & 5.0 \\
		West & 1.5 $\pm$ 1.2 & 3.0\\
     	\hline
    \end{tabular}
    \tablecomments{1. We obtain masses by simultaneously fitting three NFW profiles. 2. Mass peak significances are measured from the $2\arcmin$ aperture by dividing the convergence  by the rms map derived from our bootstrapping analysis.}
\end{table}

\subsection{Intracluster Gas Properties} \label{sec:xray_analysis}
Using a circular ($r=146\arcsec$ or $543$ kpc) aperture centered at the X-ray peak, we determined the X-ray temperature of \cluster~to be $T_X=7.97\pm0.79$~keV ($\chi^2_{red}=0.91$).
When the mass-temperature relation of \cite{2016mantz} is employed, this temperature is 
converted to $M_{500c} = 7.01^{+1.20}_{-0.84} \times 10^{14} M_{\sun}$. 
Our X-ray mass estimate is consistent with the previous results.
\cite{2018tarrio} quote $M_{500}\sim5\times10^{14} M_{\sun}$ based on their joint analysis of the {\it Planck} SZ and {\it ROSAT} X-ray data.
The recent XMM-{\it Newton} study \citep[][]{2019bulbul} reports $M_{500}=6.49_{-0.71}^{+0.81}\times10^{14} M_{\sun}$. 
These mass estimates based on X-ray data roughly agree with our WL-based result $M_{500}=6.8\pm2.4\times10^{14}~M_{\sun}$. 
However, given the clear indication of the on-going merger and invalidity of the single-halo assumption, we believe that the agreement is rather a coincidence.

Within the same $r=543$ kpc aperture, the X-ray flux of \cluster~is $f_{\rm X,0.1-2.4 keV} = 21.2 \pm 0.5 \times 10^{-13}~\rm erg~ cm^{-2} s^{-1}$, which is converted to a luminosity of $L_{\rm X,0.1-2.4 keV} = 3.41 \pm 0.10 \times 10^{44}~\rm erg~s^{-1}$.
This {\it Chandra} luminosity is in good agreement with the {\it ROSAT} result $L_{\rm X,0.1-2.4 keV} = 3.26\pm0.88\times10^{44} \rm ~erg~s^{-1}$ reported by \cite{2004bohringer}.
The relation between the total radio luminosity of the halo $P_{1.4 \rm GHz}=3.4\pm0.1 \times 10^{24} \rm~W~Hz^{-1}$ and the measured luminosity is consistent with the prediction from the $L_X - P_{1.4 \rm GHz}$ scaling relation of \cite{2012feretti}.

\begin{figure}[t]
    \includegraphics[width=1\columnwidth]{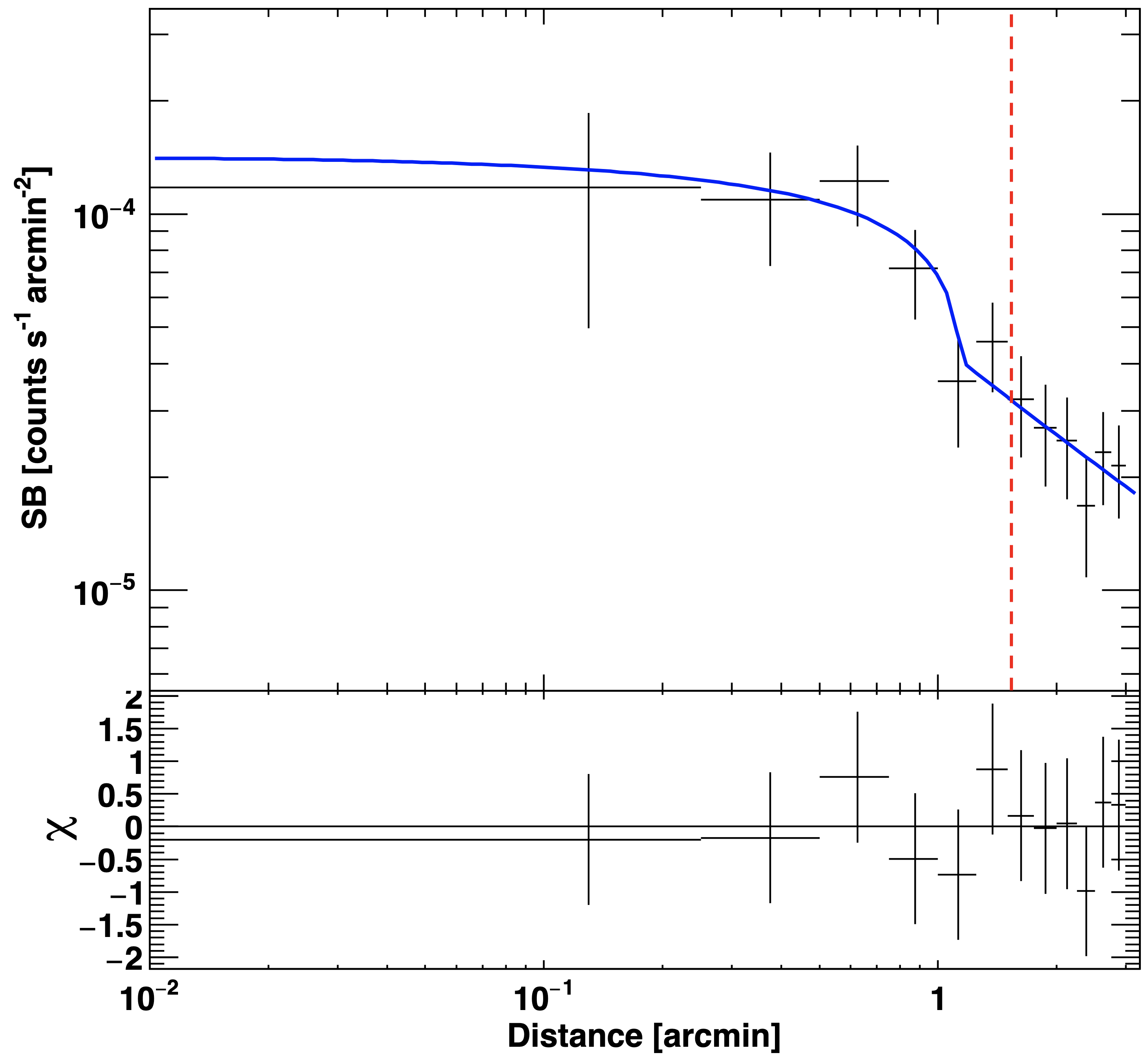}
    \caption{X-ray surface brightness profile across the relic.
    Black crosses are the data points (see 
    Figure~\ref{fig1}A for the selected regions)
    while
    the best-fit broken power-law model based
    on $\tt{PROFFIT~v1.5}$ \citep[][]{2011eckert}
    is shown in blue.  The red dashed line indicates the location of the relic boundary shown in Figure~\ref{fig1}A.
    }
    \label{fig2}
\end{figure}

Although radio relics are believed to be tracers of merger shocks, only a few clusters have shown to possess corresponding shock features in X-ray. 
Our {\it Chandra} data analysis suggests that \cluster~may belong to this rare class possessing a density jump across the relic. From the green ``panda" regions depicted in Figure~\ref{fig1}A, we determined the
density compression $\mathcal{C}=1.8\pm0.5$ as shown in Figure~\ref{fig2}. From the same regions, we measured the temperatures of the pre- and post-shock regions to be $T_X=7.3 \pm 3.3 \rm ~keV$ and $20 \pm 12 \rm ~keV$, respectively. Given the current statistics, this temperature difference is insignificant.

\section{Discussion} \label{sec:discussion}

\subsection{Too Flat Spectral Index for a Relic?}
\label{sec:spectral_index}

If particles are advected downstream and do not suffer from too strong adiabatic losses and reacceleration processes,
the integrated spectral index is steeper than the injection spectral index by $\mytilde0.5$ \citep[][]{2012bruggen, 2014brunetti}.
Thus, the implied injection spectral index $\alpha_{inj}=-0.26 \pm 0.06$ of the \cluster~relic (\textsection\ref{sec:radio_results}) 
is significantly
flatter than the theoretical upper limit $\alpha_{inj}<-0.5$ allowed in the DSA model.
According to the recent review by van Weeren et al. (2019), integrated spectral indices of the relics are within the range $-1.5 < \alpha_{int} < -1.0$. Therefore, taken at face value, the $\alpha_{int}$ value of \cluster~is unusual.
However, one must remember that we derived $\alpha_{int}$ from the narrow bandwidth (800-1088~MHz), which has yet to be verified by observations at other frequencies. Note that the spectral index of the radio relic could be biased toward a steeper value because the feature blends into the halo, which has a relatively steep spectral index. To avoid any significant contamination from the radio halo, we measured the flux density of the relic from a high resolution image (see Figure~\ref{radio_galaxy}).

In order to examine a consistency, we retrieved the archival MOST data (843 MHz), which has a beam size of $45\arcsec \times 54\arcsec$ at the location of the relic. We degraded the resolution of our ASKAP-EMU image to match this resolution and derive an aperture correction factor of 1.37 for the same polygon aperture.
With this aperture correction, we obtained a flux density of $17.9 \pm 0.3~\mbox{mJy}$, which gives a spectral index of $-0.89 \pm 0.19$, consistent with the ASKAP-EMU measurement $-0.76 \pm 0.06$.

Similarly to \cluster, flat integrated spectral indices are observed in some other clusters . For example, the radio relics of Abell 2256 \citep[][]{2012vanweeren, 2015trasatti} and Abell 3667 \citep[][]{2014hindson} have spectral indices of $\alpha_{int} = -0.83 \pm 0.03$ and $ -0.9 \pm 0.1$, respectively.

If the flatness is confirmed, it would imply that the cooling time of the downstream electrons is significantly longer than the age of the shock inferred from the merger scenario (\textsection\ref{sec:merger_scenario}).
This may happen if the relic originates from the re-acceleration of a cloud of seed relativistic plasma \citep[e.g.,][]{2012kang}. 
In this case the observed thickness of the relic traces the scale of the pre-existing plasma and the role of the plasma aging would be insignificant, provided that the shock has just crossed the cloud and that the shock crossing time is much shorter than the electron cooling time.

As mentioned in \textsection\ref{sec:radio_results}, the radio image hints at a possible presence of a ``bridge" connecting the relic to the nearby radio galaxy (Figure~\ref{radio_galaxy}).
If a future study reveals a spectral index gradient along the bridge, \cluster~might become one of the strong cases supporting the re-acceleration scenario as previously shown by A3411-12 \citep[][]{2017vanweeren, 2019andrade}. 

\subsection{Merger Scenario}
\label{sec:merger_scenario}
Both location and orientation of radio relics provide constraints on merger scenarios. 
The current \cluster~relic orientation suggests that the merger might be happening in the east-west direction, which implies that the relic is the result of the collision between the middle and eastern subclusters.
Under the assumption that the shock was generated at the impact and has been propagating to the west with nearly the same speed as the collision speed in the plane of the sky, its location (detected by the relic) can be used as an indicator of the time-since-collision (TSC).
Here, we present our estimates of the TSC in two ways. In one method, we infer the collision velocity using the timing argument \citep[][]{2002sarazin}. In the other, we use our Mach number measurements.

The timing argument (based on the assumption that the two clusters freefall to each other from an infinite separation) gives a relative velocity of $\mytilde 2,000~\mbox{km}~\mbox{s}^{-1}$ at the separation $d\sim1$~Mpc. 
If we further assume that this velocity is representative of the impact velocity, the current $\mytilde0.5$~Mpc separation of the relic from the central substructure implies a TSC of $\mytilde0.3$~Gyr.

The density compression $\mathcal{C}=1.8\pm0.5$ (\textsection\ref{sec:xray_analysis}) corresponds to the Mach number $\mathcal{M}=1.6\pm0.5$ under the Rankine-Hugoniot shock conditions.
In order to derive a collision speed, we need to compute the sound speed $c_s$, which is estimated to be $c_s\sim1,300~\mbox{km}~\mbox{s}^{-1}$ from the pre-shock temperature $\mytilde7$~keV.
The resulting collision speed is $\mytilde2,000~\mbox{km}~\mbox{s}^{-1}$, in good agreement with the value from the timing argument.

As mentioned in \textsection\ref{sec:spectral_index}, we cannot obtain an injection spectral index $\alpha_{inj}$ from the integrated spectral index $\alpha_{int}$ under the stationary shock conditions.
Assuming that $\alpha_{int}$ is a lower limit on $\alpha_{inj}$, we can convert  $\alpha_{inj}\gtrsim-0.76$ to $\mathcal{M}\gtrsim 2.9$, which in turn corresponds to TSC$\lesssim0.2$~Gyr.

In summary, although we believe that the above TSC estimates should be refined with future studies, we note that the current dataset of \cluster~indicates
that the merger shock might have originated from a recent merger ($0.2-0.3$~Gyr).

\section{Conclusions} \label{sec:conclusions}
From the deep high-resolution ASKAP-EMU pilot 300 sq. deg survey, we discovered a $\mytilde0.5$~Mpc-scale radio relic in the massive galaxy cluster SPT-CL~2023-5535 at $z=0.23$. 
We also confirmed the existence of the  $\mytilde1$~Mpc$\times0.5$~Mpc radio halo previously reported as a halo candidate.

Our study with the multi-wavelength data including {\it Chandra} and DECam shows that 1) the radio halo coincides with the intracluster gas, 2) the cluster is composed of three subclusters, and 3) across the relic there is a hint of density jump in X-ray. 
Based on these results, we suggest that the cluster is a post-merger system, where the middle and eastern subclusters might have suffered a major collision $0.2-0.3$~Gyr ago.
The cluster may belong to a rare class of radio relic clusters, where the integrated spectral indices of the relics are flatter than the test-particle DSA limit.
It possibly indicates the presence of pre-accelerated fossil electrons from the neighboring radio galaxy that are re-accelerated because of the merger shock.

There are well-known theories regarding particle acceleration by merger-induced turbulence for radio halos and shock (re-)acceleration for radio relics. 
However, the physics of these mechanisms in the ICM is still poorly known. 
Multi-wavelength observations in this framework provides a fundamental guide for the theory. With only $\mytilde60$ known radio relic systems to date, clearly one outstanding difficulty is the small sample size. Fortunately, a few giant radio surveys with state-of-the-art telescopes (e.g., SKA, LOFAR, etc.) are planned for the coming decade. The ASKAP-EMU survey, as one important program, will greatly increase the sample size by at least two orders of magnitude. The current study based on its pilot 300 sq. deg data demonstrates its tremendous potential when the full survey becomes available and supported by other multi-wavelength data.

\hfill \break
M. J. Jee acknowledges support for the current research from the National Research Foundation (NRF) of Korea under the programs 2017R1A2B2004644 and 2020R1A4A2002885. 
Partial support for LR comes from US National Science Foundation grant AST 17-14205 to the University of Minnesota. M. Yoon acknowledges support from the National Research Foundation of Korea (NRF) grant funded by the Korea government (MSIT) under no.2019R1C1C1010942.
M. Yoon acknowledge support from the Max Planck Society and the Alexander von Humboldt Foundation in the framework of the Max Planck-Humboldt Research Award endowed by the Federal Ministry of Education and Research.
MJM acknowledges the support of  the National Science Centre, Poland through the SONATA BIS grant 2018/30/E/ST9/00208.
CJR acknowledges financial support from the ERC Starting Grant ``DRANOEL'', number 714245.
The Australian SKA Pathfinder is part of the Australia Telescope National Facility which is managed by CSIRO. Operation of ASKAP is funded by the Australian Government with support from the National Collaborative Research Infrastructure Strategy. Establishment of the Murchison Radio-astronomy Observatory was funded by the Australian Government and the Government of Western Australia. ASKAP uses advanced supercomputing resources at the Pawsey Supercomputing Centre. We acknowledge the Wajarri Yamatji people as the traditional owners of the Observatory site.

\end{document}